\documentclass{PoS}
\usepackage{amsmath}

\title{Asymmetries for neutral pion photoproduction in the threshold region}

\ShortTitle{Asymmetries for neutral pion photoproduction in the threshold region}

\author{\speaker{David Hornidge} \\
       Mount Allison University \\
       E-mail: \email{dhornidge@mta.ca}}

\abstract{
We report on two pion-photoproduction measurements in the threshold region
conducted in the A2 collaboration at MAMI with the almost 4$\pi$ Crystal Ball
detector. The first was with a linearly polarized photon beam and unpolarized
liquid-hydrogen target.  The data analysis is now complete and the linearly
polarized beam asymmetry along with differential cross sections provide the
most stringent test to date of the predictions of Chiral Perturbation Theory
and its energy region of convergence.  More recently, a measurement was
performed using both circularly polarized photons and a transversely polarized
butanol frozen-spin target, with the goal of extracting both the target and
beam-target asymmetries.  From these we intend to extract $\pi N$ scattering
sensitive information for the first time in photo-pion reactions.  This will
be used to test isospin conservation and further test dynamics of chiral
symmetry breaking in QCD as calculated at low energies by Chiral Perturbation
Theory.
}

\FullConference{
	The 7th International Workshop on Chiral Dynamics \\
	August 6--10, 2012 \\
	Jefferson Lab, Newport News, Virginia, USA
}

\begin{document}

\section{Introduction}

Over an extended period of time the efforts of the A2 collaboration at Mainz
have been focused on accurate measurements of low-energy $\gamma N$ Compton
scattering and pion production reactions to perform tests of Chiral
Perturbation Theory (ChPT) predictions.  Study of the $\gamma p \rightarrow
\pi^0 p$ reaction started with the original MAMI accelerator and a small
detector to observe the $\pi^0 \rightarrow \gamma \gamma$ decay~\cite{Beck},
and then followed with increasingly more accurate experiments to obtain the
relatively small cross section~\cite{Fuchs,Schmidt}.  A parallel effort was
also carried out at Saskatoon~\cite{Sask} during this period.  The Mainz work
has been building up to the sensitive spin observables~\cite{Schmidt}, and the
present generation photo-pion production experiments that we are concentrating
on include accurate measurements of the cross sections, polarized photon
asymmetries, and polarized target and beam-target asymmetries $T$ and $F$.
These experiments have been carried out with circularly and linearly
polarized, tagged photons and with the almost $4\pi$ Crystal Ball and TAPS
detector system.  They provide very stringent tests of dynamical
models~\cite{DMT2001} and predictions based on chiral symmetry breaking in
QCD\@.
 
\section{Photon Asymmetry in $\vec{\gamma}p\to\pi^0p$}

\begin{figure}[ht]
\centerline{\includegraphics[width=0.4\textwidth]{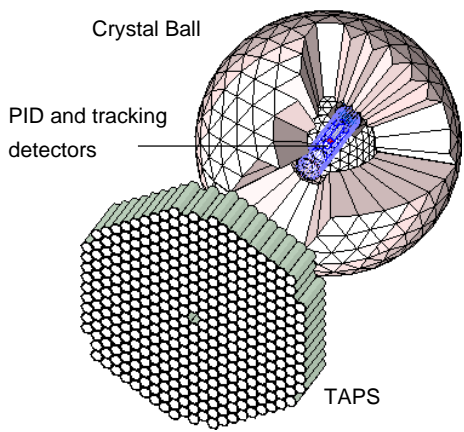}}
\caption{A cut-away view of the CB-TAPS detector system.  The solid-angle
coverage is approximately 96\% of $4\pi$.}
\label{fig:CB-TAPS}
\end{figure}

In December 2008 we performed an investigation of the $\vec{\gamma}p
\rightarrow \pi^0p$ reaction with a linear polarized photon beam and a liquid
H$_2$ target using the Glasgow-Mainz photon tagger and the CB-TAPS detector
system in the A2 hall at MAMI\@.  The purpose was to execute the most accurate
measurement to date of the differential cross section from threshold through
the $\Delta$ region, and to greatly improve our previous polarized photon
asymmetry measurement~\cite{Schmidt}.  Note that the detector set-up in our
original measurement covered only about 30\% of $4\pi$, meaning that the
detection efficiency for the two-photon channel of $\pi^0$ decay was on the
order of 10\%.  The more recent experiment made use of the CB-TAPS set-up
shown in Figure~\ref{fig:CB-TAPS}---covering $\approx 96\%$ of
$4\pi$---resulting in a detection efficiency for the $\pi^0$ channel of
roughly 90\%.  This fact alone made for a large improvement in the accuracy
and counting rates for the new measurement.
In addition, a higher energy electron beam resulted in a significant increase
in the degree of polarization for the incident photon beam.  The other main
difference between the old and new measurements is that sufficient empty
target data was taken for the latter, which turned out to be crucial due to
the contribution to the asymmetry from the $0^+$ nuclei in the kapton target
windows.  Due to poor statistics in the older TAPS experiment, the polarized
photon asymmetry, $\Sigma$, was integrated over the entire incident photon
energy range, leading to data only at the cross section weighted energy
average of 159.5 MeV\@.

\begin{figure}[ht]
\begin{center}
\includegraphics[height=0.38\textheight]{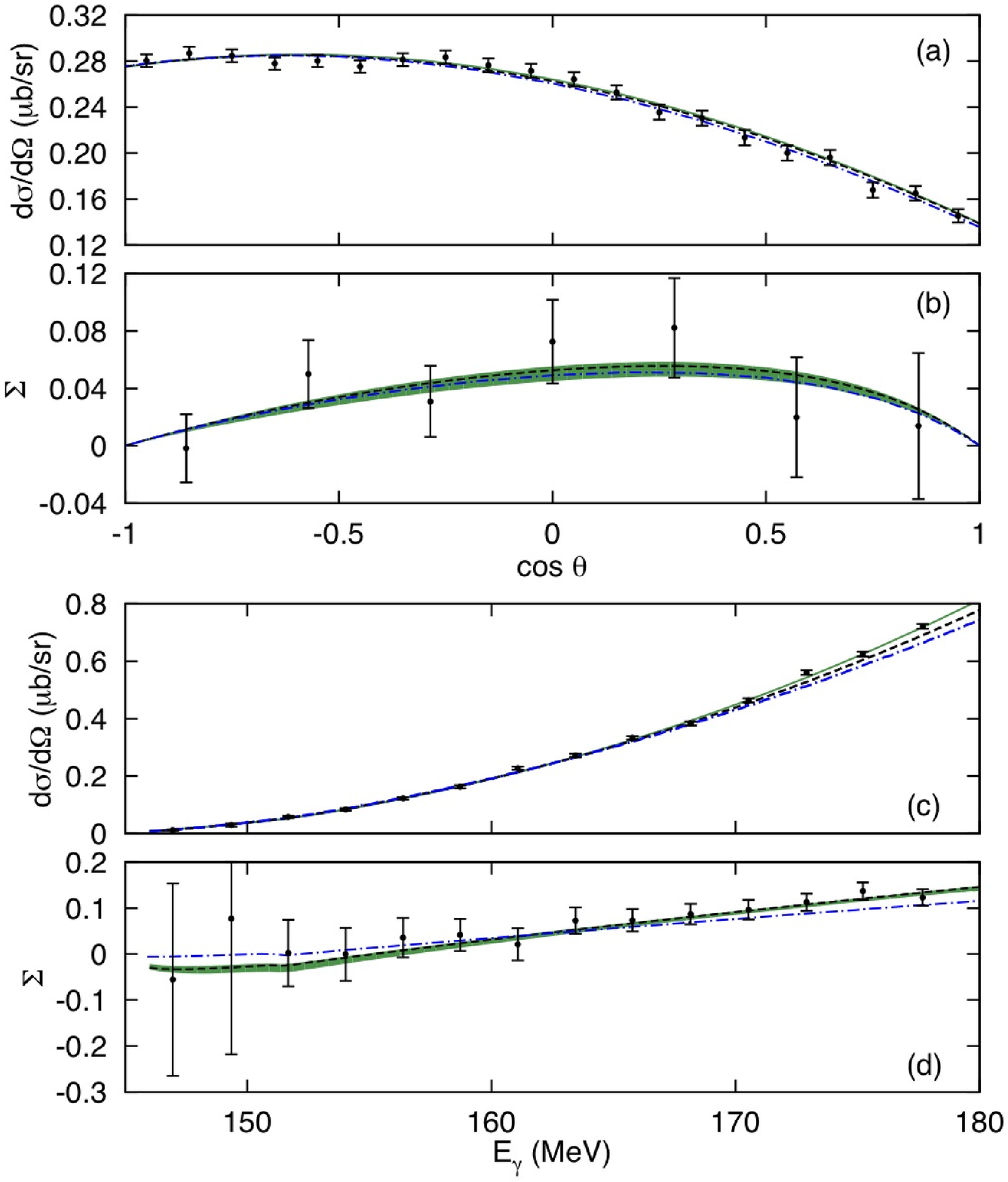}
\hfill
\includegraphics[height=0.38\textheight]{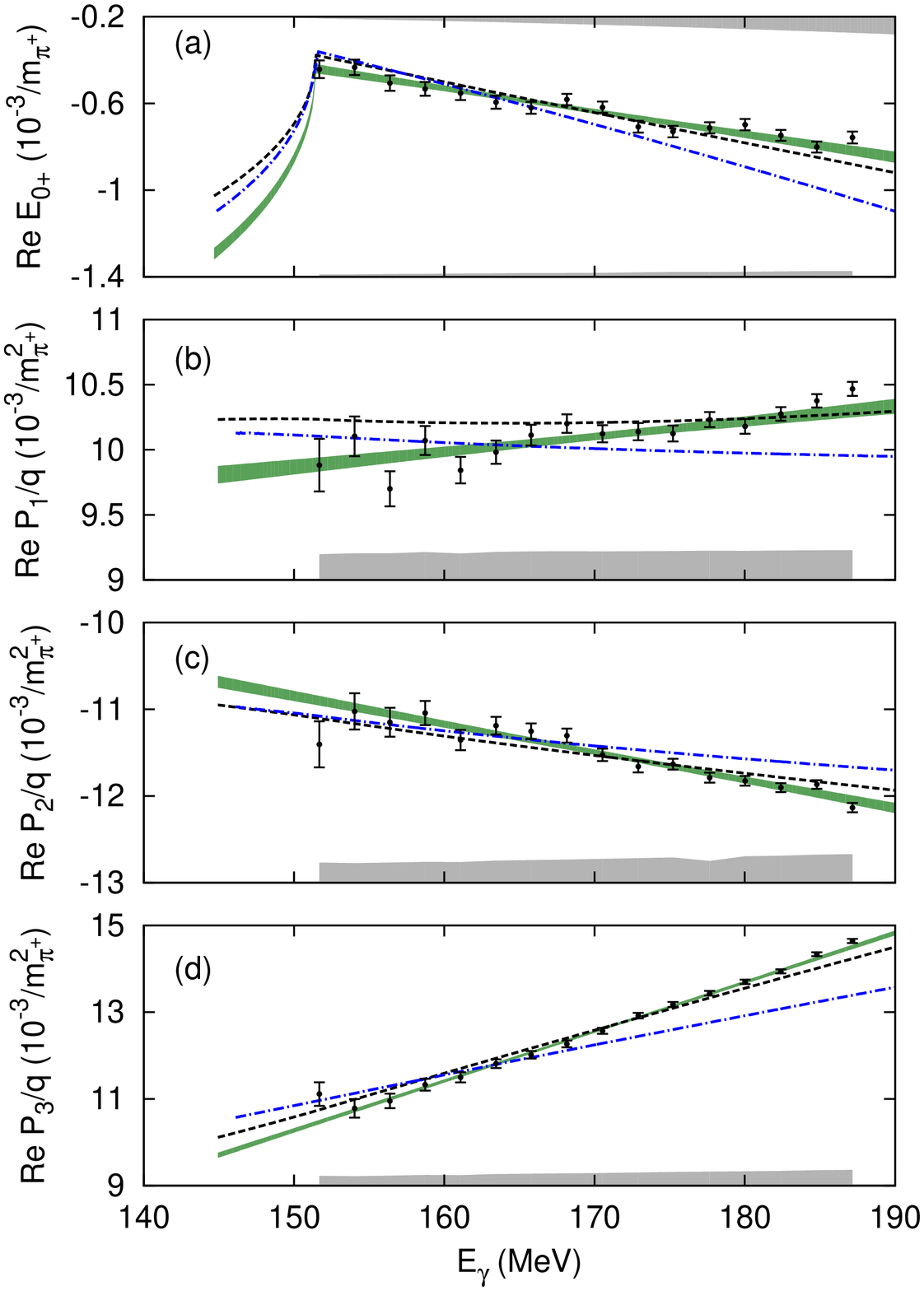}
\end{center}
\caption{Left column:  Differential cross sections in $\mu$b/sr (a) and photon
asymmetries (b) for $\pi^0$-production as a function of pion CM production
angle $\theta$ for an incident photon energy of $163.4\pm1.2$~MeV\@.  Energy
dependence of the differential cross sections (c) and photon asymmetries (d)
at $\theta \simeq 90\pm3^\circ$.  Errors shown are statistical only, without
the systematic uncertainty of 4\% for $d\sigma/d\Omega$ and 5\% for $\Sigma$.
The theory curves are dashed (black) HBChPT~\cite{Cesar-ChPT}, dash-dotted
(blue) relativistic ChPT~\cite{Hilt,Hilt2}, and (green) an empirical fit with
an error band.  Note that in (c) and (d) the two points in incident photon
energy below $\pi^+$ threshold are included.  Right column: Empirical
multipoles as a function of incident photon energy: (a) $ReE_{0^+}$, (b)
$ReP_1/q$, (c) $ReP_2/q$, (d) $ReP_3/q$.  The points are single-energy fits to
the real parts of the $S$- and $P$-wave multipoles, and the empirical fits
from (\protect\ref{eq:swave}) and (\protect\ref{eq:pwaves}) are shown with
(green) statistical error bands.  The $\pm$ systematic uncertainty for the
single-energy extraction is represented as the gray area above the energy
axis, and the systematic uncertainty in the $S$-wave extraction due to the
uncertainty in the size of the $D$-wave contributions is given by the gray
area at the top of (a).  The theory curves are the same as in the left
column.}
\label{fig:dxs_mult}
\end{figure}

For the new measurement, the data analysis is finished and sample results for
the differential cross section and photon asymmetry are shown in the left
column of Figure~\ref{fig:dxs_mult} (more details can be found in
Ref.~\cite{Dave-PRL}).  The top two panels of the left column are at an
incident-photon energy of 163.4\,MeV as a function of pion CM production angle
$\theta$, and the bottom two are at $\theta \simeq 90\pm3^\circ$ as a function
of incident photon energy.  We have photon asymmetries from just above
threshold in 2.4-MeV-wide bins, and differential cross sections from threshold
into the $\Delta$ region.  Fitting of the data has been done for the
low-energy constants in both HBChPT~\cite{Cesar-ChPT} (solid curves in
Figure~\ref{fig:dxs_mult}) and relativistic ChPT~\cite{Hilt,Hilt2} (dashed
curves in Figure~\ref{fig:dxs_mult}).  Moreover, with the use of an empirical
model-independent partial-wave analysis, one can extract various coefficients
from the differential cross sections and photon beam asymmetry, and
comparisons can be made between the extracted coefficients and the theory
predictions (green band in Figure~\ref{fig:dxs_mult} with error explained
below).  The $S$- and $P$-wave multipoles then appear only in the coefficients
allowing for a direct comparison of theory and experiment.

Specifically, the differential cross section and photon asymmetry can be
expanded in terms of the pion center-of-mass (CM) angle, $\theta$, in the
following way
\begin{equation*}
\frac{d\sigma}{d\Omega}(\theta) = \frac{q}{k}
	\left(a_0 + a_1\cos\theta + a_2\cos^2\theta\right)
	\qquad \mathrm{and} \qquad
	\Sigma(\theta) = \frac{q}{k}\left(b_0\sin^2\theta\right)
	/\frac{d\sigma}{d\Omega}(\theta)
\end{equation*}
where $q$ is the CM momentum of the outgoing pion, $k$ is the CM momentum of
the incident photon, and $a_0$, $a_1$, $a_2$, and $b_0$ are the coefficients
containing the $S$- and $P$-waves.  In terms of the multipoles, these
coefficients are given by
\begin{align*}
a_0 &= |E_{0^+}|^2 + P_{23}^2
& P_1 &= 3E_{1^+} + M_{1^+} - M_{1^-} \\
a_1 &= 2ReE_{0^+}P_1
& P_2 &= 3E_{1^+} - M_{1^+} + M_{1^-} \\
a_2 &= P_1^2 - P_{23}^2
& P_3 &= 2M_{1^+} + M_{1^-} \\
b_0 &= \frac{1}{2}\left(P_3^2 - P_2^2\right)
& P_{23}^2 &= \frac{1}{2} (P_2^2 + P_3^2)
\end{align*}
Note that the $D$-waves make a contribution \emph{even in the threshold
region}, but the current data set did not have enough precision to extract
them; they have been taken from the Born terms.

For the model-independent partial-wave analysis, the empirical fits to the
data start with the following ansatz for the multipoles
\begin{align}
E_{0^+}(W) &= E_{0^+}^{(0)} + E_{0^+}^{(1)}
	\left(\frac{E_\gamma-E_\gamma^{\mathrm{thr}}}{m_{\pi^+}}\right)
	+ i\beta\frac{q_{\pi^+}}{m_{\pi^+}}, \label{eq:swave} \\
P_{i}(W) &= \frac{q}{m_{\pi^+}} \left[P_{i}^{(0)}
	+ P_{i}^{(1)}\left(
	\frac{E_\gamma-E_\gamma^{\mathrm{thr}}}{m_{\pi^+}}
	\right)\right], \label{eq:pwaves}
\end{align}
where here $E_\gamma$ and $E_\gamma^{\mathrm{thr}}$ are in the lab frame, and
$E_{0^+}^{(0)}, E_{0^+}^{(1)}, P_{i}^{(0)}, P_{i}^{(1)}$ (with $i = 1,2,3$)
are constants that are fit to the data.


Based on unitarity, the cusp parameter in (\ref{eq:swave}) has the value
$\beta = Re E_{0^+}(\gamma p \rightarrow \pi^+n)a_\mathrm{cex}(\pi^+n \rightarrow
\pi^0 p)$~\cite{AB-lq}.  Using the best experimental value of
$a_\mathrm{cex}(\pi^{-} p \to \pi^{0} n) = -(0.122 \pm 0.002) /m_{\pi^+}$
obtained from the observed width in the $1s$ state of pionic
hydrogen~\cite{Gotta}, assuming isospin is a good symmetry, i.e.\
$a_\mathrm{cex}(\pi^{+} n \rightarrow \pi^{0} p) = -a_\mathrm{cex}(\pi^{-} p
\rightarrow \pi^{0} n)$, and the latest measurement for $E_{0+}({\gamma} p \to
\pi^{+} n) = \left( 28.06 \pm 0.27 \pm 0.45 \right)\times
10^{-3}/m_{\pi^+}$~\cite{Korkmaz}, we obtain $\beta = \left( 3.43 \pm 0.08
\right) \times 10^{-3}/m_{\pi^+}$, which was employed in the empirical fit.
The uncertainty introduced by the errors in $\beta$ and isospin breaking are
smaller than the statistical uncertainties of the multipole extraction
depicted in Fig.~\ref{fig:dxs_mult}.

The extracted multipoles are displayed in the right column of
Figure~\ref{fig:dxs_mult} along with the theoretical calculations.  The
points are single-energy fits to the real parts of the $S$- and $P$-wave
multipoles, and the energy-dependent fits from (\ref{eq:swave}) and
(\ref{eq:pwaves}) are shown with the error band.  The imaginary part of the
$S$-wave multipole $E_{0^+}$ was taken from unitarity (\ref{eq:swave}) with
the value of the cusp parameter explained above, the imaginary parts of the
$P$-waves were assumed to be negligible, and the $D$-wave multipoles were
calculated in the Born approximation.  The impact of $D$-waves in the $P$-wave
extraction is negligible~\cite{Cesar-D} but in the $S$-wave it can
be sizeable.  In order to assess the uncertainties in the $S$-wave extraction
associated to our $D$-wave prescription we have estimated the uncertainty from
the difference between the Born terms and DMT dynamical model
in~\cite{DMT2001}.  This error estimation is depicted in
Figure~\ref{fig:dxs_mult} as a gray area at the top of the first plot.
As was the case for the observables, there is very good agreement between the
two ChPT calculations and the empirical values of the multipoles for energies
up to $\simeq160$~MeV with the same pattern of deviations above that.

In conclusion, the combination of the photon asymmetry and improved accuracy
in the differential cross section has allowed us to extract the real parts of
the $S$-wave and all three $P$-wave multipoles as a function of photon energy
for the first time.  As was the case for the observables, the ChPT
calculations agree with the multipoles up to an energy of $\simeq$ 165 to 175
MeV, with the relativistic calculations deviating at the lower-energy end and
the HBChPT calculations closer to the upper end.

\section{Target and Beam-Target Asymmetries in $\vec{\gamma}\vec{p}\to\pi^0p$}

\begin{figure}[ht]
\begin{center}
\includegraphics[width=0.49\textwidth]{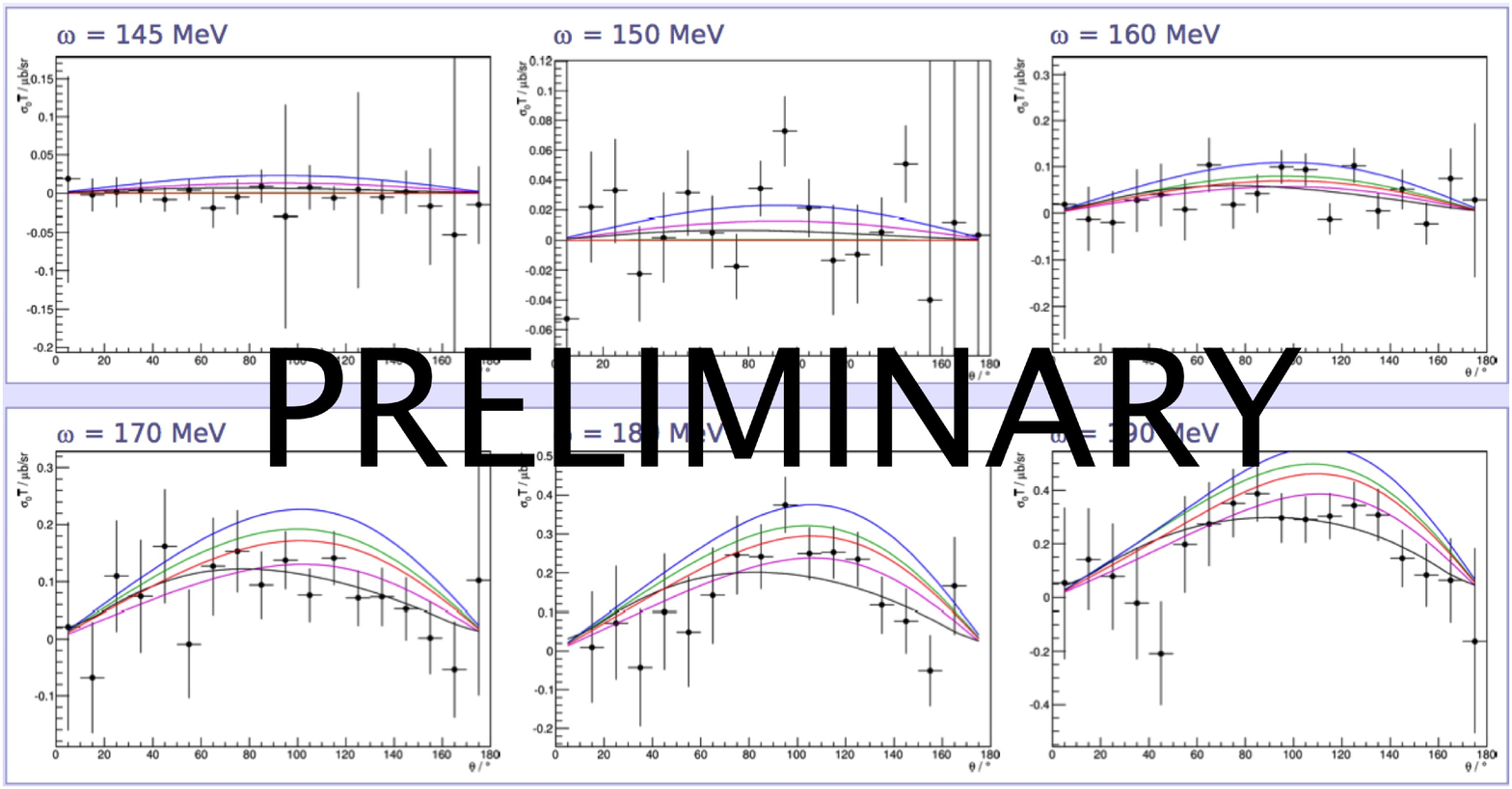}
\hfill
\includegraphics[width=0.49\textwidth]{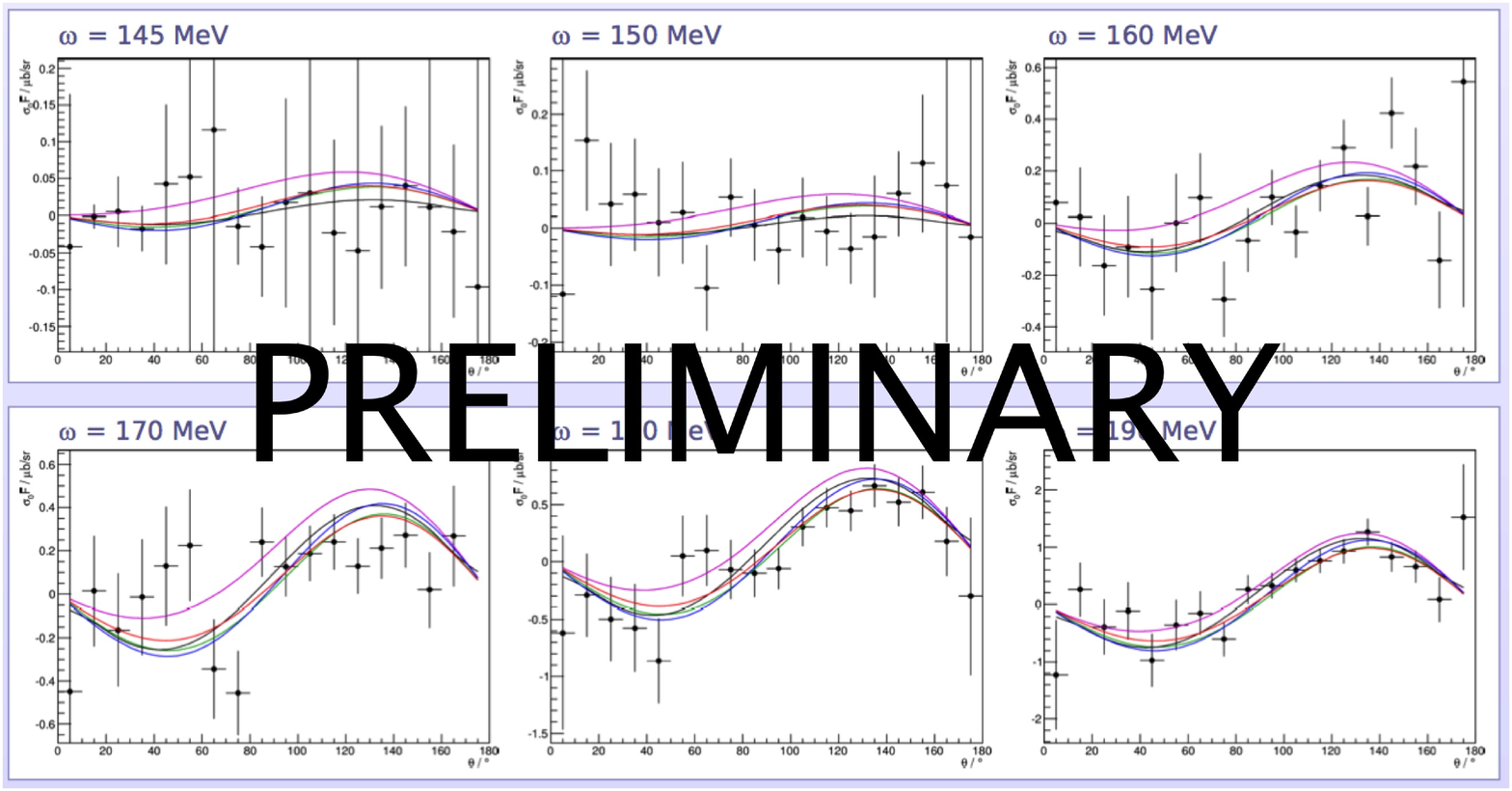} \\[1.5ex]
\includegraphics[width=0.49\textwidth]{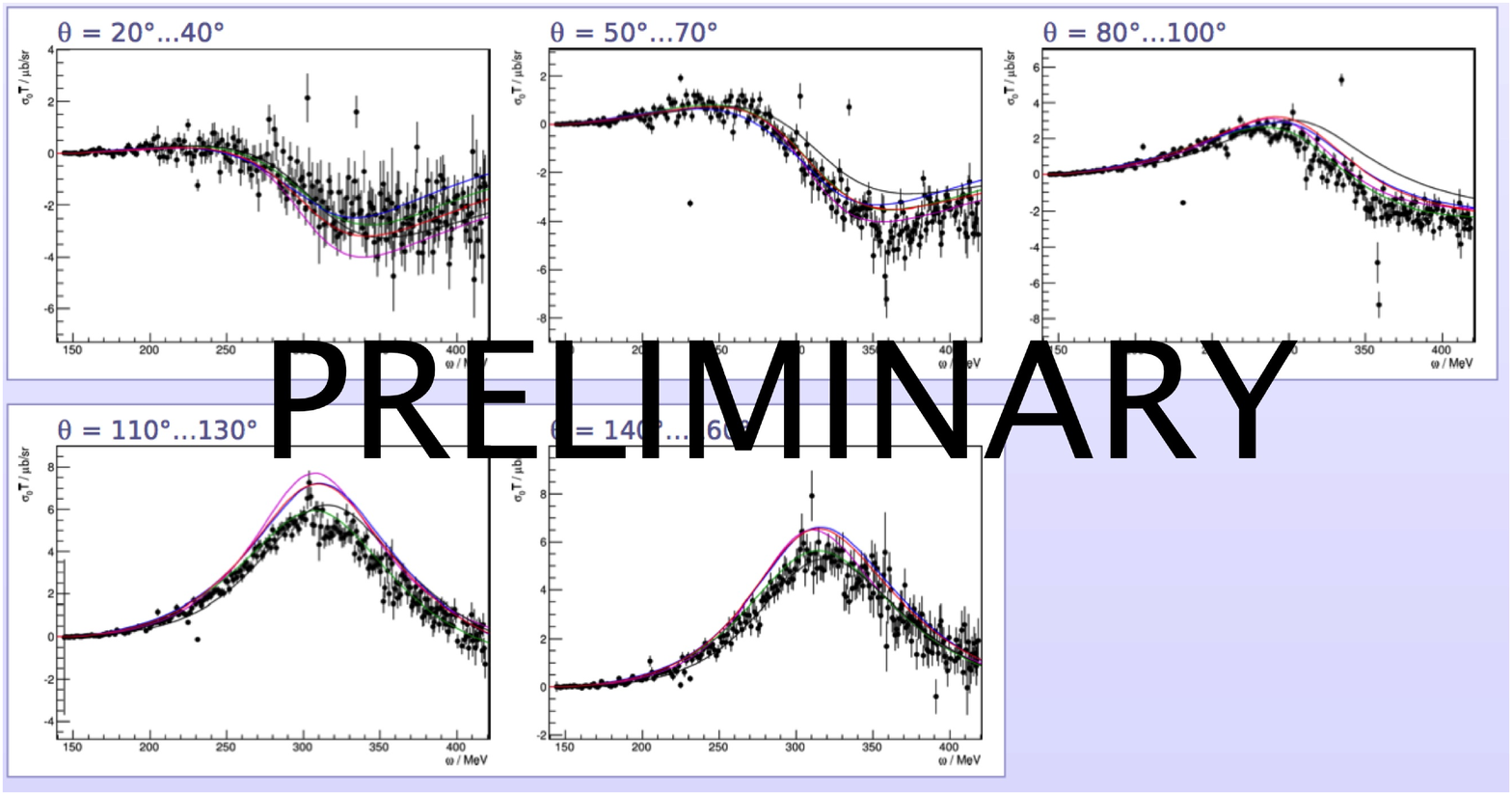}
\hfill
\includegraphics[width=0.49\textwidth]{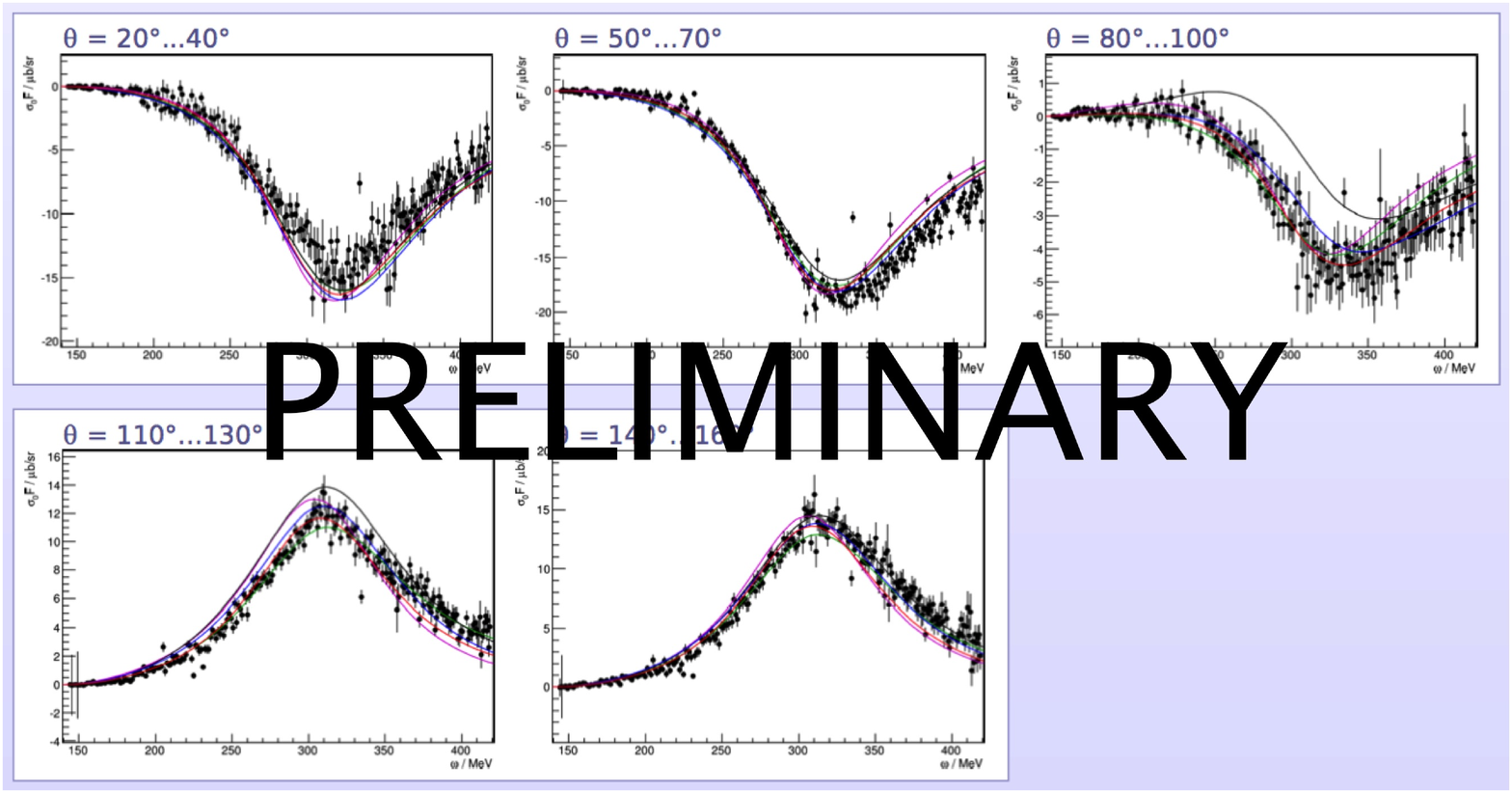}
\end{center}
\caption{Preliminary target asymmetries $T$ (left panels) and $F$ (right
panels) at six incident photon energies as a function of pion CM production
angle (top panels), and for various $\theta$ bins as a function of incident
photon energy (bottom panels).  The errors are statistical and the lines are
predictions of the DMT~\cite{DMT2001} (green), MAID~\cite{MAID} (red), SAID~\cite{SAID} (blue),
Bonn-Gatchina~\cite{BnGa} (black), and Gie\ss
en~\cite{Giessen} (purple).}
\label{fig:TF}
\end{figure}

We have performed a precise measurement of the $\vec{\gamma}\vec{p} \to
\pi^0p$ reaction from threshold to the $\Delta$ resonance using a circularly
polarized photon beam and a transverse polarized target~\cite{A2-prop} to
obtain the polarized target asymmetry, $T$, and the double polarization
observable $F$. Note that the latter is sensitive to the $D$-wave multipoles
that have recently been shown to be important in the near threshold
region~\cite{Cesar-D}.  Preliminary asymmetries for $T$ and $F$ are shown in
Figure~\ref{fig:TF} as both a function of angle and energy compared to
numerous models predictions~\cite{DMT2001,MAID,SAID,BnGa,Giessen}.

The target asymmetry, $T$, has the following form
\begin{equation*}
T = ImE_{0^+}^{\pi^0p}(P_3-P_2)\sin\theta
\end{equation*}
which should allow us to make a direct determination of the imaginary part of
the $S$-wave amplitude, $ImE_{0^+}^{\pi^0p}$, above the $\pi^+ n$ threshold.
From this, we intend to extract the cusp parameter, $\beta$, from
(\ref{eq:swave}) and then use the known value of $E_{0^+}^{\pi^+n}$ to find
$a_{ex}(\pi^+ n\to \pi^0 p)$.  If isospin is conserved, then
\begin{equation*}
a_{ex}(\pi^+ n\to\pi^0p) = a_{ex}(\pi^-p\to\pi^0n).
\end{equation*}
At the present time the right-hand side has been measured in pionic hydrogen
with an error of $\simeq$ 1.5\%~\cite{Gotta}, and it is anticipated that
future work will reduce the uncertainty.  Any deviations from the isospin
conserving limit will test isospin breaking due to the electromagnetic
interaction and the strong interaction due to the mass difference of the up
and down quarks predicted in ChPT~\cite{Martin}.  Observation of $T$ for the
first time in the intermediate-energy region, combined with the other accurate
data which we are obtaining, will provide us with information about the $\pi
N$ phase shifts for charge states ($\pi^0p,\pi^+n$) that are not accessible to
conventional $\pi N$ scattering experiments.  This will enable us to test
isospin conservation~\cite{AB-IS}.  In addition these measurements will test
detailed predictions of Chiral Perturbation Theory~\cite{loop} and its energy
region of convergence.


\begin{thebibliography}{99}

	\bibitem{Beck} R.~Beck et al., Phys.\ Rev.\ Lett.\ \textbf{65}, (1990)
	1841--1844.
	
	\bibitem{Fuchs} M.~Fuchs et al., Phys.\ Lett.\ B \textbf{368}, (1996)
	20--25.
	
	\bibitem{Schmidt} A.~Schmidt et al., Phys.\ Rev.\ Lett. \textbf{87}, (2001)
	232501 1--4.
		
	\bibitem{Sask} J.~C.~Bergstrom et al., Phys.\ Rev.\ C \textbf{53}, (1996)
	R1052--R1056; Phys.\ Rev.\ C \textbf{55}, (1997) 2016--2023.

	\bibitem{DMT2001} S.~S.~Kamalov, S.~N.~Yang, D.~Drechsel, and L.~Tiator,
	Phys.\ Rev.\ Lett.\ \textbf{83}, (1999) 4494--4497; Phys.\ Rev.\ C
	\textbf{64}, (2001) 032201 1--5.

	\bibitem{Dave-PRL} D. Hornidge et al., Phys.\ Rev.\ Lett.\ \textbf{111},
	(2013) 062004 1--5.  D.~Hornidge and A.~.M.~Bernstein, Eur.\ Phys.\ J
	Special Topics \textbf{198}, (2011) 133--140.

	\bibitem{Cesar-ChPT} C.~{Fern{\'a}ndez-Ram{\'i}rez} and A.~M.~Bernstein,
	manuscript in preparation.

	\bibitem{Hilt} M.~Hilt, private communication.

	\bibitem{Hilt2} M.~Hilt, S.~Scherer, and L.~Tiator, manuscript in
	preparation.

	\bibitem{AB-lq} A.~M.~Bernstein, Phys.\ Lett.\ B \textbf{442}, (1998)
	20--27.

	\bibitem{Korkmaz} E. Korkmaz et al., Phys.\ Rev.\ Lett. \textbf{83}, (1999)
	3609.

	\bibitem{Cesar-D} C.~Fern\'{a}ndez Ram\'{i}rez, A.~M.~Bernstein, and
	T.~W.~Donnelly, Phys.\ Lett.\ B \textbf{679}, (2009) 41--44.
	C.~Fern\'{a}ndez Ram\'{i}rez, A.~M.~Bernstein, and T.~W.~Donnelly, Phys.\
	Rev.\ C \textbf{80}, (2009) 065201 1--15.

	\bibitem{A2-prop} M.~Ostrick, D.~Hornidge, W.~Deconinck, and A. M.~Bernstein,
	spokespersons, MAMI proposal A2-10/09, (2009).

	\bibitem{MAID} D.~Drechsel, O.~Hanstein, S.~S.~Kamalov, and L.~Tiator,
	Nucl.\ Phys.\ A \textbf{645}, (1999) 145--174; Eur.\ Phys.\ J. A
	\textbf{34}, (2007) 69--97.

	\bibitem{SAID} R.~Arndt, W.~Briscoe, I.~Strakovsky, and R.~Workman, Phys.\
	Rev.\ C \textbf{74}, (2006) 045205. 

	\bibitem{BnGa} A.~V.~Sarantsev et al., Phys.\ Lett.\ B \textbf{659}, (2008) 94.

	\bibitem{Giessen} V.~Shklyar, U.~Mosel, and H.~Lenske, Phys.\ Lett.\ B
	\textbf{650}, (2007) 172.

	\bibitem{Gotta} D.~Gotta et al., AIP Conf.\ Proc.\ \textbf{1037}, (2008)
	162--177.

	\bibitem{Martin} M.~Hoferichter, B.~Kubis, and U.-G.~Mei\ss ner, Phys.\
	Lett.\ B \textbf{678}, (2009) 65--71; Nucl.\ Phys.\ A \textbf{833}, (2010)
	18--103

	\bibitem{AB-IS} A.~M.~Bernstein, $\pi N$ Newsletter No.\ {\bf 11}, (1995),
	and article in preparation

	\bibitem{loop} V.~Bernard, N.~Kaiser, and U.-G.~Mei{\ss}ner,  Nucl.\ Phys.\
	B \textbf{383}, (1992) 442; Int.\ J. Mod.\ Phys.\ E \textbf{4}, (1995)
	193--344; Phys.\ Rev.\ Lett.\ \textbf{74}, (1995) 3752--3755; Eur.\ Phys.\
	J. A \textbf{11}, (2001) 209--216.

\end{thebibliography}
\end{document}